\documentstyle[12pt,aaspp4]{article} 
\begin{document}
\title{Is There an Age of the Universe Problem After the {\it Hipparcos} 
Data?}
\author{Murat \"Ozer}
\affil{ Department of Physics, College of Science, King Saud University
 	         P.O.Box 2455, Riyadh 11451, Saudi Arabia
\footnote{E-mail: mozer@ksu.edu.sa}}
\newcommand{\etal}{{\it et al. }}

\begin{abstract}
We have reanalyzed the age of the universe problem under the assumption 
that the lower limit on the age of the globular clusters is $11Gyr$, as 
predicted by the recent {\it Hipparcos} data. We find that the globular 
cluster and the expansion ages  in a standard $\lambda=0$ universe are 
consistent only if the present value $H_0$ of the Hubble constant is 
$\leq 60kms^{-1}Mpc^{-1}$. If $H_0>60kms^{-1}Mpc^{-1}$ some kind of 
modification of the standard $\lambda=0$ model is required. Invoking a 
(time-independent) cosmological term $\lambda$ in the Einstein field 
equations, as has been done frequently before, we have found that due to 
the gravitational lensing restrictions a flat universe with the present 
matter density parameter $\Omega_M<0.5$ is not problem-free. A nonflat 
universe with $\Omega_M\leq 1$ does not suffer from the age problem if 
$H_0\leq 75kms^{-1}Mpc^{-1}$.
\end{abstract}

%{\bf Key words:} Cosmology: theory,  gravitational lensing.\\

\section{INTRODUCTION}

A lower limit on the present age $t_0$ of the universe is determined by 
estimating the age of the oldest objects in our galaxy, the globular 
clusters
 (hereafter GC). These are stellar systems that contain about $10^5$ stars
in the halo surrounding the galactic disk. The key element in estimating 
the age of a typical GC is the determination of its distance from us. To 
this end, the primary observational technique is main-sequence fitting 
against subdwarfs with well known parallaxes. The distance obtained this 
way or otherwise is used to convert the measured apparent magnitude of a 
GC to the absolute magnitude. The age is then estimated by applying a 
stellar evolution model. The estimates obtained by different astronomers 
agree rather well. For example, Bolte \& Hogan (1995) find $15.8\pm 2.1Gyr$, 
Chaboyer \etal (1996)
find $14.6\pm 1.7Gyr$, and Sandquist \etal (1996) find $13.5\pm 1Gyr$.  
These time scales are to be compared with the expansion age of the universe 
predicted by the
standard model of cosmology (hereafter SM)
\footnote{Felten \& Isaacman(1986) call the models
with $\lambda=0$ "standard models". However, we follow the general trend in
the literature and call the totality of them  the "standard model" and refer 
to each case by its $k$ value (see, for example, Misner, Thorne, \& Wheeler 
1970; Weinberg 1972.)  The SM with $k=0$ is called the Einstein-de Sitter 
model (Felten \& Isaacman 1986).}
which requires the knowledge of the present value $H_0$ of the Hubble 
constant. Even though the estimates of $t_0$ from GCs are based on the 
stellar evolution madels, which are essentially the same, the situation is 
not the same for the $H_0$ estimates.
There are a number of different techniques (see the review by Trimble 1996)
which give values that differ substantially from each other. We present the 
most quoted estimates: $H_0=50-55kms^{-1}Mpc^{-1}$ (Tammann \& Sandage 1996) 
and $H_0=73\pm 10 kms^{-1}Mpc^{-1}$ (Freedman, Madore, \& Kennicut 1997)
\footnote{Reid (1997) has argued that this Freedman \etal (1997) value of 
$H_0$ is reduced to $H_0=68\pm 9kms^{-1}Mpc^{-1}$ because the 
{\it Hipparcos} data reveal a 7\% increase in the distances inferred from 
the previous ground-based data.}
. In a SM flat universe $t_0$ would be $13Gyr$ and $8.2Gyr$ if $H_0$ were 
$50kms^{-1}Mpc^{-1}$ and $80kms^{-1}Mpc^{-1}$, respectively. Whereas in a 
SM open universe with $\Omega_M=0.1$, $\Omega_M$ being the present 
nonrelativistic matter density parameter, the ages would be $17.6Gyr$ and 
$11Gyr$ for the same $H_0$ values as above. Thus researchers were rightfully
led to think that if $H_0$ has as large a value as determined by Freedman 
\etal (1997) then the expansion age and the GC age of the universe are in 
conflict with each other.

An immediate solution to this so called age of the universe problem was 
suggested by including a time-independent cosmological constant $\lambda$ 
in the Einstein field equations (Peebles 1984; Blome \& Priester 1985; 
Klapdor \& Grotz 1986). The gravitational lensing studies, however, have 
shown that the cosmological constant cannot be as large as one desires to 
increase the expansion age to the level of GC age lest too many lensing 
events are predicted (Kochanek 1993, 1995; Maoz \& Rix 1993). Recently, the 
supernova magnitude-redshift approach (Perlmutter \etal 1997) has given 
$\Omega_{\Lambda}<0.51$ (95\% confidence level) for a flat universe which 
is significantly lower than the gravitational upper limit 
$\Omega_{\Lambda}<0.66$ of Kochanek (1995). Thus it had been concluded that 
the apparent contradiction between the GC age and the axpansion age could 
not be reconciled in a flat universe by invoking a time-independent 
cosmological constant. This was the status of the age of the universe 
problem before {\it Hipparcos}. The purpose of this paper is to reexamine 
this problem in the light of the lower limit of $11Gyr$ on the GC age put by
the {\it Hipparcos} data (Reid 1997; Feast \& Catchpole 1997; see also the 
news report by Schwarzschild (1997)).

\section{THE AGE OF THE UNIVERSE PROBLEM}

The relation between the present value $H_0$ of the Hubble constant 
$H=\dot a/a$, where $a$ is the scale factor of the universe and 
$\dot a=da/dt$, and the present age $t_0$ is given by (Al-Rawaf \& Taha 
1996)
\footnote{Equations (1a) and (1c) agree numerically with those given in 
Weinberg (1972) which uses a different but equivalent functional form.}
\begin{mathletters} %eq.1a,b,c
\begin{eqnarray}
H_{0}t_{0} &=& 
\frac{1}{(1-\Omega_{M})}\left[1-\frac{\Omega_{M}}{(1-\Omega_{M})^{1/2}}
sinh^{-1}(\Omega_{M}^{-1}-1)^{1/2}\right]\hspace{2mm},\hspace{2mm}k=-1\\
&=& 2/3 \hspace{2mm}, \hspace{8.5cm} k=0 \\
&=& \frac{1}{(\Omega_{M}-1)}\left[\frac{\Omega_{M}}{(\Omega_{M}-1)^{1/2}}
sin^{-1}(1-\Omega_{M}^{-1})^{1/2}-1\right] \hspace{2mm},\hspace{2mm}k=1.
\end{eqnarray}
\end{mathletters}
Here $\Omega_M$ is the present value of the nonrelativistic matter density
parameter defined as the ratio of the present nonrelativistic matter density
to the present critical energy density
\begin{equation}%eq.(2)
\Omega_M=\frac{\rho_M}{\rho_c}=\frac{\rho_M}{3H_0^2/8\pi G}.
\end{equation}
Expressing the Hubble constant as $H_0=100hkms^{-1}Mpc^{-1}$, the age in 
billion years is given by $t_0(Gyr)=9.78(H_0t_0)/h$, where $(H_0t_0)$ is 
given in eq.(1). In Figure 1, we depict  $t_0$ against $\Omega_M$ and $h$ 
in the SM. It is seen that $t_0$ is below the {\it Hipparcos} lower limit of
$11 Gyr$ for large values of $h$. Thus it can be stated safely that the age 
of the universe problem still survives if $h$ is large.

In Table 1, we display the maximum values of $h$ for which $t_0=11Gyr$ 
against $\Omega_M$.  Note that the maximum $h$ values in Table 1 almost 
fall in the lower and upper limits of Freedman \etal (1997). Thus for each 
$\Omega_M$, if $h$ is greater than those given in Table 1, there is an age 
problem. For example, if $\Omega_M=0.5$ and $h>0.67$ or $\Omega_M=1$ and 
$h>0.593$ the age problem survives. Now the problem is, however, milder in 
the sense that before {\it Hipparcos} the age problem was thought to exist 
even for moderate values of $h$ whereas it now exists for large values of 
$h$. Emphatically, the SM has no age problem if $h<0.593\approx 0.6$.

Supposing that  there is an age problem, one line of attack,  as in the 
pre {\it Hipparcos} era, is to invoke a (time-independent) cosmological 
constant $\lambda$ in the Einstein field equations (Peebles 1984; Blome 
\& Priester 1985; Klapdor \& Grotz 1986)
\begin{equation}% eq.(3)
R_{\mu\nu}-\frac{1}{2}g_{\mu\nu}R-\lambda g_{\mu\nu}=-8\pi GT_{\mu\nu},
\end{equation}
where $R=R_{\alpha}^{\alpha}$ and $T_{\mu\nu}$ is the energy-momentum 
tensor. For a homogeneous and isotropic universe described by the 
Robertson-Walker metric
\begin{equation}%eq.(4)
ds^{2}=-dt^{2}+a(t)^{2}\left[\frac{dr^{2}}{1-kr^{2}}+r^{2}(d\theta^{2}+
sin^{2}\theta d\phi^{2})\right]
\end{equation}
the energy-momentum is assumed to have the perfect fluid form
\begin{equation}%eq.(5)
T_{\mu\nu}=diag(\rho, p, p, p),
\end{equation}
where $p$ is the pressue of the matter described by $\rho$. Equations (3) 
and (4) give (with c, the speed of light, set to 1)
\begin{equation} %eq.(6)
H^2=\left(\frac{\dot a}{a}\right)^{2}=\frac{8 \pi G}{3}\rho(t)+
\frac{\lambda}{3}-\frac{k}{a^{2}},
\label{eqn:friedmann1}
\end{equation}
where $k=-1, 0, 1$ for a spatially open, flat and  closed universe, 
respectively. At present, the universe is believed to be dominated by 
nonrelativistic massive matter rather than relativistic matter (radiation). 
It proves to be very usefull to define the current cosmological constant 
density parameter 
\begin{equation} %eq.(7)
\Omega_{\Lambda}=\frac{\rho_{\Lambda}}{\rho_c}=\frac{\lambda/8\pi G}
{\rho_c}=\frac{\lambda}{3H_0^2},
\end{equation}
and the current curvature density parameter
\begin{equation} %eq.(8)
\Omega_k=-\frac{\rho_k}{\rho_c}=-\frac{k}{H_0^2a_0^2},
\end{equation}
where $a_0$ is the current value of the scale factor $a$ of the universe. 
When written in terms of the present values equation (6) gives the 
constraint
\begin{equation} %eq.(9)
\Omega_M+\Omega_{\Lambda}+\Omega_k=1
\end{equation}
Equations (3) and (5) under (4) give the energy conservation equation in 
the matter dominated era
\begin{equation} %eq.(10)
d[\rho_M(t)a^3+\frac{\lambda}{8\pi G}a^3]+[p_M(t)-\frac{\lambda}{8\pi G}]
da^3=0,
\end{equation}
where the pressure $p_M$ of nonrelativistic matter is negligible. Thus it 
follows from eq.(10) that $\rho_M(t)=\rho_M a_0^3/a^3$ and the relation 
between $H_0$ and $t_0$ is
\begin{equation} %eq.(11)
H_0t_0=\int_0^1 y^{1/2}[\Omega_M(1-y)+\Omega_{\Lambda}(y^3-y)+y]^{-1/2}dy,
\end{equation}
where  $\Omega_k$ has been eliminated by using eq.(9). Now a flat universe 
with $\Omega_M<1$ is rendered possible by postulating the existance of the 
cosmological term $\lambda$ such that $\Omega_M+\Omega_{\Lambda}=1$. The 
value of $k$ not fixed {\it a priori}, a numerical investigation of eq.(11) 
reveals that it is always possible to find a set of three parameters
$(\Omega_M, \Omega_{\Lambda}, h_{max})$ for which $t_0=11Gyr$.

However, the achievement of a cosmological constant to solve the age problem 
and to have a flat universe with $\Omega_M<1$ may be illusory. The magnitude 
of $\Omega_{\Lambda}$ required to solve the age problem may turn out to be 
too large to predict plausible number of gravitational lensing events. 
Therefore, each such set of parameters $(\Omega_M, \Omega_{\Lambda}, 
h_{max})$ need to be confronted with the gravitational lensing statistics, 
which we address ourselves next.

\section{THE GRAVITATIONAL LENSING STATISTICS}

The integrated 
probability, the so-called optical depth, for lensing by a population of 
singular isothermal spheres of constant comoving density relative to the 
Einstein-de Sitter model, is 
\begin{equation} %eq.(12)
P_{lens}=\frac{15}{4}\left[1-\frac{1}{(1+z_s)^{1/2}}\right]^{-3}
\int_0^{z_s}\frac{(1+z)^2}{E(z)}\left[\frac{d(0, z)d(z, z_s)}{d(0, z_s)}
\right]^2dz
\label{eqn:plens}
\end{equation}
(Carroll, Press \& Turner 1992)
where 
\begin{equation} %eq.(13)
E(z)^2=(1+z)^2(1+z\Omega_{M})-z(z+2)\Omega_{\Lambda}
\label{eqn:e(z)}
\end{equation}
and is defined by
\begin{equation} %eq.(14)
\left(\frac{\dot a}{a}\right)^2=H_0^2E(z)^2
\end{equation}
(Peebles 1993). Note that $P_{lens}=1$ for the Einstein-de-Sitter model 
(in which $\Omega_k=0$, $\Omega_{M}=1$ and $\Omega_{\Lambda}=0$)
. $z=(a_0/a)-1$ is the redshift and $z_s$ is the redshift of the source 
(quasar). The angular diameter distance from redshift $z_1$ to redshift 
$z_2$ is 
\begin{equation}%eq.(15)
d(z_1,z_2)=\frac{1}{(1+z_2)\mid\Omega_{k}\mid^{1/2}}
sinn\left[\mid\Omega_{k}\mid^{1/2}\int_{z_1}^{z_2}\frac{dz}{E(z)}\right]
\label{eqn:d}
\end{equation}
where  "$sinn$" is defined as $sinh$ if
 $\Omega_{k}>0$, as $sin$ if $\Omega_{k}<0$ and as unity if $\Omega_{k}=0$ 
in which case the $\mid\Omega_{k0}\mid^{1/2}$'s disappear from 
eq.(\ref{eqn:d}).
To determine how much of $P_{lens}$ is permissible, we refer to the work of 
the Supernova Cosmology Project (Perlmutter \etal 1997). Using the initial 
seven of more than 28 supernovae discovered, Perlmutter \etal (1997) have 
recently measured $\Omega_M$ and $\Omega_{\Lambda}$. For $\Omega_M<1$, they 
find $\Omega_{\Lambda}<0.51$ at the 95\% confidence level for a flat 
universe, and $\Omega_{\Lambda}<1.1$ for the more general case 
$\Omega_M+\Omega_{\Lambda}$ unconstrained
\footnote{But of course $\Omega_M+\Omega_{\Lambda}+\Omega_k=1$}. 
In Table 2 we present $P_{lens}$ against $\Omega_M$ and $\Omega_{\Lambda}$ 
for a typical source redshift of $z_s$=2.
Table 2 helps us determine the maximum allowed value of $P_{lens}$. It is 
seen that for $\Omega_{\Lambda}=0.5$, which is the maximum allowed value 
according to Perlmutter \etal (1997), the corresponding $P_{lens}$ is 1.92. 
Thus we shall assume that $P_{lens}$ cannot be much larger than 2. Having 
determined the upper limit on $P_{lens}$, we depict in Table 3 the three 
parameters $\Omega_M$, $\Omega_{\Lambda}$ and $h_{max}$ in a flat universe 
and the corresponding gravitational lensing prediction for $t_0=11Gyr$. In 
preaparing Table 3, we have first fixed $\Omega_M$ and calculated $H_0t_0$ 
from eq.(11) with $\Omega_{\Lambda}=1-\Omega_M$, and finally obtained the 
maximum value of $h$ from $h_{max}=9.78(H_0t_0)/11$. 

Discarding those set of parameters which yield $P_{lens}>2$ or have 
$\Omega_{\Lambda}>0.5$, first we confirm, from Table 3, the previous 
conclusions that a cosmological constant cannot solve the age problem in a 
flat universe with $\Omega_M<0.5$ due to too many lensing predictions. 
Next, we see that the maximum allowed value of $h$ in a flat universe is 
about 0.74-0.75. This is to be compared with the pre {\it Hipparcos} lower 
limits for the age. For $t_0=13$ and $14Gyr$ the $h_{max}$ values are 0.64 
and 0.60 in a flat universe, respectively.

As for a nonflat universe, one may either fix $\Omega_M$ and 
$\Omega_{\Lambda}$ first and then calculate $h_{max}$ to give $t_0=11Gyr$, 
or one may fix $\Omega_M$ and $h_{max}$ first and then calculate the
$\Omega_{\Lambda}$ value from eq.(11) by trial and error to give again 
$t_0=11Gyr$. We have chosen the second option and constructed Figures 2 
and 3, which are contour diagrams of $h_{max}$ (for $t_0=11 Gyr$) in the 
$(\Omega_M, \Omega_{\Lambda})$ and $(\Omega_M, P_{lens})$ planes.

It is seen that for each contour there is a minimum value of $\Omega_M$ 
before which the  age is greater or equal to $11 Gyr$ for 
$\Omega_{\Lambda}=0$. In drawing Figures 2 and 3, we have assumed that the 
maximum allowed value of $\Omega_{\Lambda}$ is about 1.1, in accordance 
with the findings of Perlmutter \etal (1997). The age problem is seen to 
survive for $\Omega_M\geq 0.3$ only if $h$ is as large as 0.8 for which 
lensing predictions are larger than 2. There is no age problem in a 
non-flat universe provided $h\leq 0.75$ for all $\Omega_M\leq 1$.

\section{CONCLUSIONS}

That the {\it Hipparcos} data (Reid 1997; Feast \& Catchpole 1997) imply 
that GCs may be as young as $11Gyr$ has raised the hopes to reconcile the 
age of GCs and the expansion age of the universe. We have studied this 
matter in this work. As is well known, and as born out by our results, the 
realization of this hope depends solely on what the value of $H_0$ is. If 
$H_0$ is as large as the upper limit of the Freedman \etal (1997) value, 
the age of the universe problem continues to exist in the SM. The problem, 
however, is now milder than it was before {\it Hipparcos}. Previously, it 
was thought to exist even for moderate values of $h$, whereas it seems to 
exist for large values of $h$ now. If, however, $H_0$ is as low as favored 
by Tammann \& Sandage (1996) then the GC and the expansion ages of the 
universe are consistent with each other in the SM.

Assuming that $H_0$ is high and hence modifying the SM by invoking a 
(time-independent) cosmological term in the Einstein field equations, as 
has been done before (Peebles 1984; Blome \& Priester 1985; 
Klapdor \& Grotz 1986), we have  
confirmed the  conclusion of previous workers that due to lensing 
restrictions the age problem still survives in a flat universe for 
$\Omega_M<0.5$, and at the same time concluded that $h$ cannot be larger 
than about 0.75. As for a nonflat universe, we have shown that the age 
problem does not exist for all $\Omega_M\leq 1$ provided $h\leq 0.75$. 

 The above mentioned hope is realized in the SM only if $h\leq 0.6$ 
(see Table 1). Otherwise, some kind of modification of the SM is called for. 
One such, and the most-studied, attempt is the inclusion of a cosmological 
term in the field equations. With such a term, the age problem has a better 
standing in a nonflat (open or closed) universe with $\Omega_M\leq 1$. It 
should be noted, in the light of recent works, that such a cosmological term
need not be a pure time-independent constant. Scalar fields, cosmic strings 
or some kind of stable textures with an energy density varying as $a^{-2}$ 
lead to viable cosmological models that stand as alternatives to the SM 
(Kamionkowski \& Toumbas 1996; Spergel \& Pen 1997; \"Ozer 1999).

\clearpage
%FIGURE CAPTIONS:
%
%Figure 1: The age of the universe in the SM for $k=-1$ 
%(solid lines), $k=0$ (dots) and $k=1$ (dashed lines) versus the present 
%value of the matter density parameter $\Omega_M$.
%
%Figure 2: Contours of $h_{max}$ for which $t_0=11Gyr$ in the 
%$(\Omega_M, \Omega_{\Lambda}$) plane.
%
%Figure 3: Contours of $h_{max}$ for which $t_0=11Gyr$ in the 
%$(\Omega_M, P_{lens})$ plane.
%
%\clearpage
%TABLE CAPTIONS:
%Table 1: Maximum values of $h$ in the SM for which $t_0=11Gyr$.
%Table 2: Normalized optical depths. 
%Table 3: Maximum values of $h$ in a flat universe for which $t_0=11Gyr$.
%
\clearpage
\begin{deluxetable}{ccccccccccc}
\tablecaption{Maximum values of $h$ in the SM for which 
$t_0=11Gyr\tablenotemark{a}$}
\label{table:1}
\tablewidth{0pt}
\tablehead{
\colhead{$\Omega_{M}$} &\colhead{0.1} &\colhead{0.2 } &\colhead{0.3} 
&\colhead{0.4} &\colhead{0.5} &\colhead{0.6} &\colhead{0.7} &\colhead{0.8}
&\colhead{0.9} &\colhead{1.0}}
\startdata
  $h_{max}$ &0.799 &0.753 &0.719 &0.692 &0.67 &0.651 &0.634 &0.619 &0.605 
&0.593\nl
\tablenotetext{a}{Note that if $h=h_{max}$ then $t_0=11Gyr$, if $h>h_{max}$ 
then $t_0<11Gyr$ and if $h<h_{max}$ then $t_0>11Gyr$.}
\enddata
\end{deluxetable}
%
%\clearpage
\begin{deluxetable}{ccc}
\tablecaption{Normalized optical depths. }
\label{table:2}
\tablewidth{0pt}
\tablehead{
\colhead{$\Omega_{M}$} &\colhead{$\Omega_{\Lambda}$} &\colhead{$P_{lens}$} }
\startdata
  0 &1.0 &13.25\nl
0.1 &0.9 &5.98\nl
0.2 &0.8 &3.94\nl
0.3 &0.7 &2.93\nl
0.4 &0.6 &2.33\nl
0.5 &0.5 &1.92\nl
0.6 &0.4 &1.63\nl
0.7 &0.3 &1.42\nl
0.8 &0.2 &1.25\nl
0.9 &0.1 &1.11\nl
1.0 &0   &1.00\nl
1.0 &1.1 &1.61\nl
0.8 &1.1 &1.99\nl
0.6 &1.1 &2.57\nl
0.4 &1.1 &3.61\nl
0.2 &1.1 &6.05\nl
\enddata
\end{deluxetable}
%
%\clearpage
\begin{deluxetable}{cccc}
\tablecaption{Maximum values of $h$ in a flat universe for which $t_0=11Gyr$. }
\label{table:3}
\tablewidth{0pt}
\tablehead{
\colhead{$\Omega_{M}$} &\colhead{$\Omega_{\Lambda}$} &\colhead{$h_{max}$} 
&\colhead{$P_{lens}\tablenotemark{a}$} }
\startdata
0.1 &0.9 &1.14 &5.98\nl
0.2 &0.8 &0.96 &3.94\nl
0.3 &0.7 &0.86 &2.93\nl
0.4 &0.6 &0.79 &2.33\nl
0.45 &0.55 &0.76 &2.11\nl
0.5 &0.5 &0.74 &1.92\nl
0.6 &0.4 &0.70 &1.63\nl
0.7 &0.3 &0.67 &1.42\nl
0.8 &0.2 & 0.64 &1.25\nl
0.9 &0.1 &0.61 &1.11\nl
1.0 &0   &0.59 &1.00\nl
\enddata
\tablenotetext{a}{Recall that $P_{lens}$ is independent of $h$ 
(see equations (12)-(15)).}
\end{deluxetable}
\begin{figure}
\epsscale{.8}
\plotone{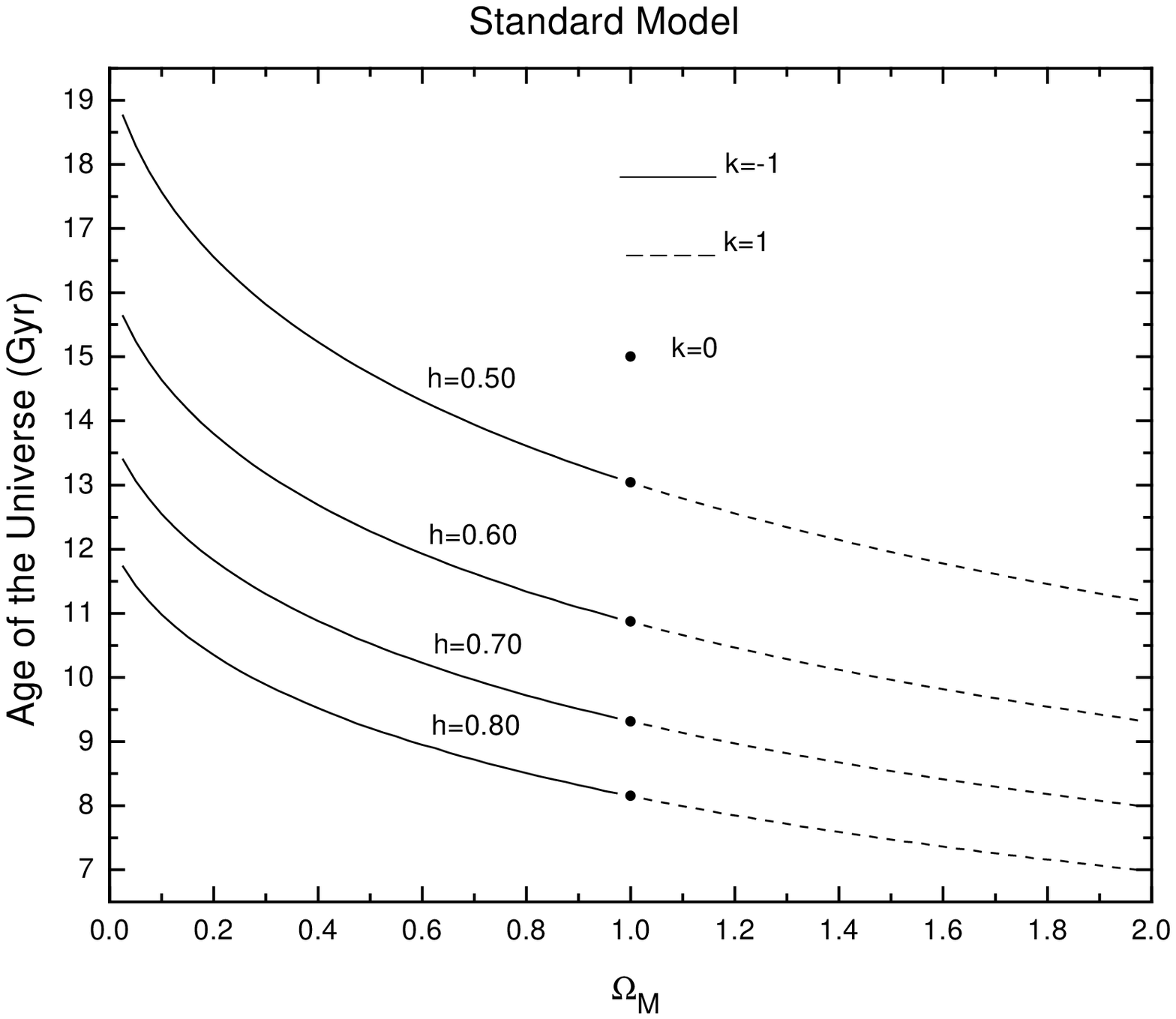}
\caption{The age of the universe in the SM for $k=-1$ 
(solid lines), $k=0$ (dots) and $k=1$ (dashed lines) versus the present 
value of the matter density parameter $\Omega_M$.}
\label{figure1}
\end{figure}
\begin{figure}
\epsscale{.8}
\plotone{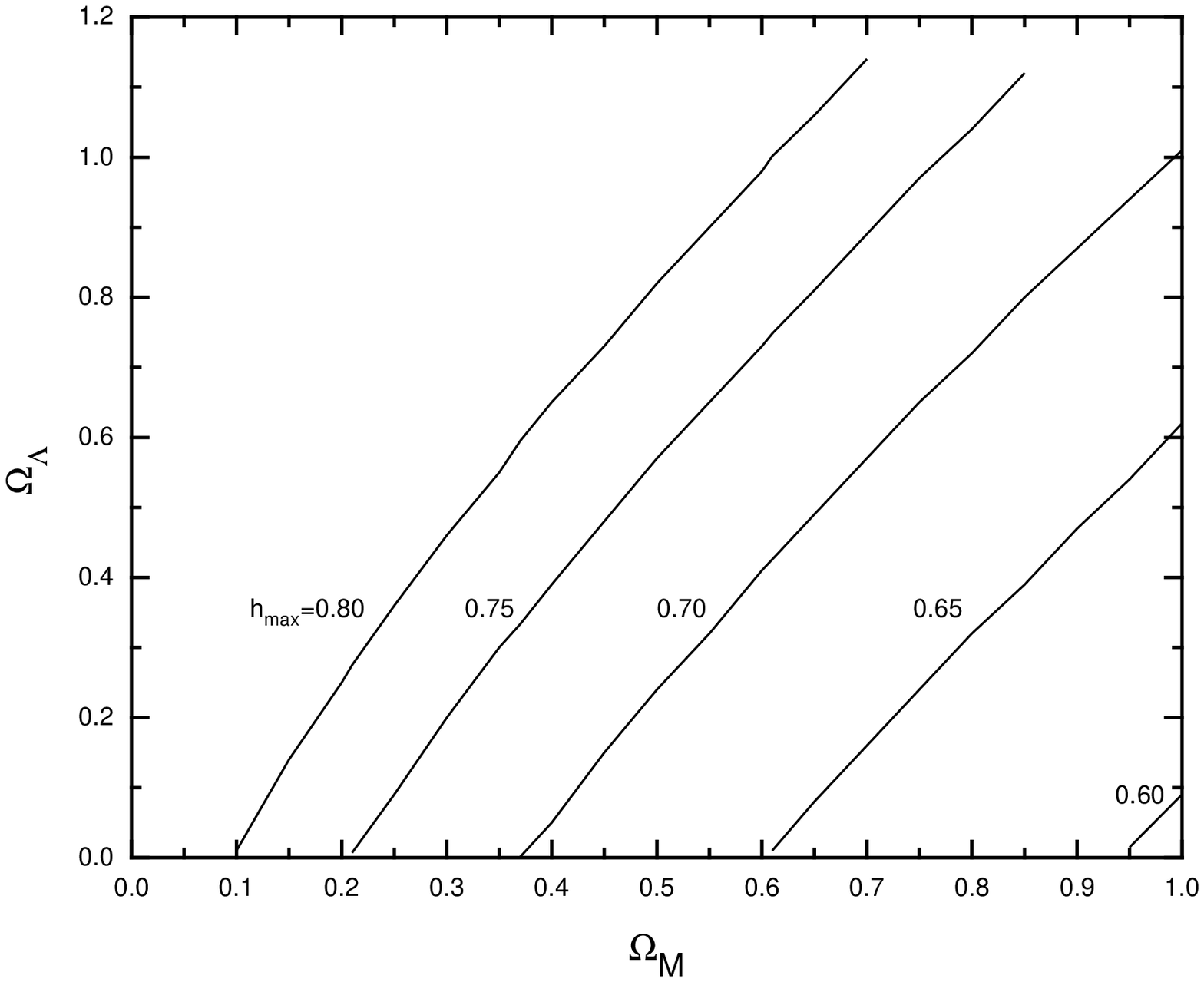}
\caption{Contours of $h_{max}$ for which $t_0=11Gyr$ in the 
$(\Omega_M, \Omega_{\Lambda}$) plane.}
\label{figure2}
\end{figure}
\begin{figure}
\epsscale{.8}
\plotone{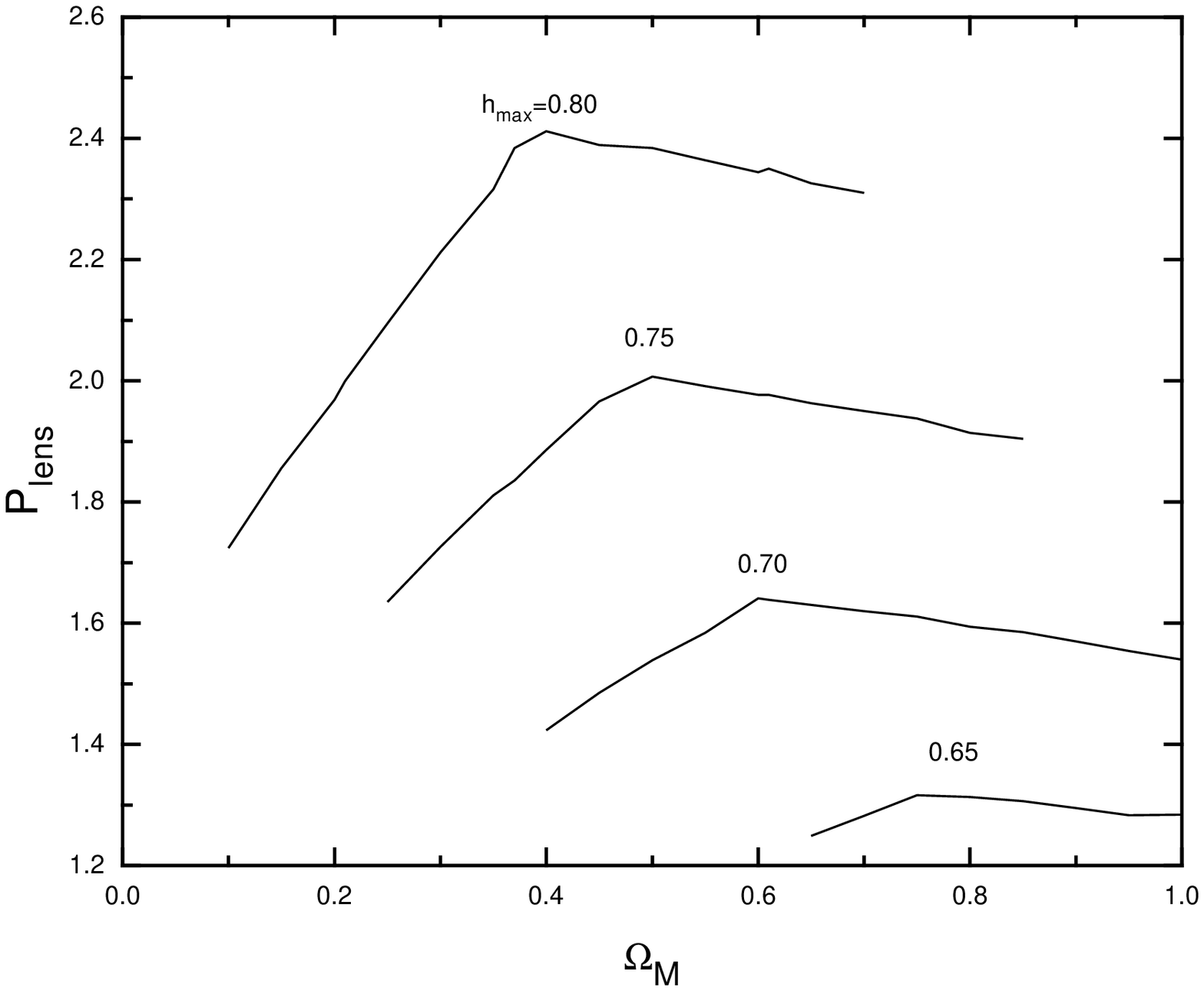}
\caption{Contours of $h_{max}$ for which $t_0=11Gyr$ in the 
$(\Omega_M, P_{lens})$ plane.}
\label{figure3}
\end{figure}

\end{document}